\documentclass[letterpaper, 10 pt, conference]{ieeeconf}  

\IEEEoverridecommandlockouts                             
\let\labelindent\relax
\usepackage[colorlinks,linkcolor=blue]{hyperref}
\usepackage{amsmath, amssymb, mathrsfs}
\usepackage[utf8]{inputenc}
\usepackage[T1]{fontenc}
\usepackage{graphicx}
\usepackage[table]{xcolor}
\usepackage{colortbl}
\usepackage{tabularx}
\usepackage{enumitem}
\usepackage{array, xcolor, colortbl}
\usepackage{multirow}
\usepackage{tikz}
\usepackage{algorithm}
\usepackage{algpseudocode}
\usepackage{booktabs} 
\definecolor{green_arrow}{RGB}{0, 128, 0}
\definecolor{red_arrow}{RGB}{255, 0, 0}
\definecolor{lightgray}{gray}{0.9}
\definecolor{lightblue}{rgb}{0.75, 0.85, 0.95}
\definecolor{lightgreen}{rgb}{0.75, 0.95, 0.75}
\definecolor{lightyellow}{rgb}{0.95, 0.95, 0.75}

\title{\LARGE \bf
Detection of Peri-Pancreatic Edema using Deep Learning and Radiomics Techniques}

\author{Ziliang Hong\textsuperscript{3}, Debesh Jha\textsuperscript{1}, Koushik Biswas\textsuperscript{1}, Zheyuan Zhang\textsuperscript{1}, Yury Velichko\textsuperscript{1}, Cemal Yazici\textsuperscript{3}, Temel Tirkes\textsuperscript{4}, \\
Amir Borhani\textsuperscript{1}, Baris Turkbey\textsuperscript{2}, Alpay Medetalibeyoglu\textsuperscript{1}, Gorkem Durak\textsuperscript{1},  Ulas Bagci\textsuperscript{1}
\thanks{*This work is supported by the NIH funding: R01-CA246704, R01-CA240639, U01-DK127384-02S1, and U01-CA268808.}
\thanks{$^{1}$D. Jha, K. Biswas, A. Bourhani, G. Durak, Z. Zhang, and U. Bagci are with Machine \& Hybrid Intelligence Lab, Northwestern University. $^{2}$B. Turkbey is with National Cancer Institute. $^{3}$Z. Hong and C. Yazici are with University of Illionis at Chicago. $^{4}$T. Tirkes is with Indiana University.
}}

\begin{document}
\maketitle
\thispagestyle{empty}
\pagestyle{empty}
\begin{abstract}
Pancreatitis is a major public health issue worldwide; studies show an increase in the number of people experiencing pancreatitis. Identifying peri-pancreatic edema is a pivotal indicator for identifying disease progression and prognosis, emphasizing the critical need for accurate detection and assessment in pancreatitis diagnosis and management. This study \textit{introduces a novel CT dataset sourced from 255 patients with pancreatic diseases, featuring annotated pancreas segmentation masks and corresponding diagnostic labels for peri-pancreatic edema condition}.  With the novel dataset, we first evaluate the efficacy of the \textit{LinTransUNet} model, a linear Transformer based segmentation algorithm, to segment the pancreas accurately from CT imaging data. Then, we use segmented pancreas regions with two distinctive machine learning classifiers to identify existence of peri-pancreatic edema: deep learning-based models and a radiomics-based eXtreme Gradient Boosting (XGBoost). The LinTransUNet achieved promising results, with a dice coefficient of 80.85\%, and mIoU of 68.73\%. Among the nine benchmarked classification models for peri-pancreatic edema detection, \textit{Swin-Tiny} transformer model demonstrated the highest recall of $98.85 \pm 0.42$ and precision of $98.38\pm 0.17$. Comparatively, the radiomics-based XGBoost model achieved an accuracy of $79.61\pm4.04$  and recall of $91.05\pm3.28$, showcasing its potential as a supplementary diagnostic tool given its rapid processing speed and reduced training time. To our knowledge, \textit{this is the first study aiming to detect peri-pancreatic edema} automatically. We propose to use modern deep learning architectures and radiomics together and created a benchmarking for the first time for this particular problem, impacting clinical evaluation of pancreatitis, specifically detecting peri-pancreatic edema. Our code is available \url{https://github.com/NUBagciLab/Peri-Pancreatic-Edema-Detection}.
\end{abstract}
\begin{figure*}[!t]
\centering

    \centerline{\includegraphics[width= 0.8\textwidth]{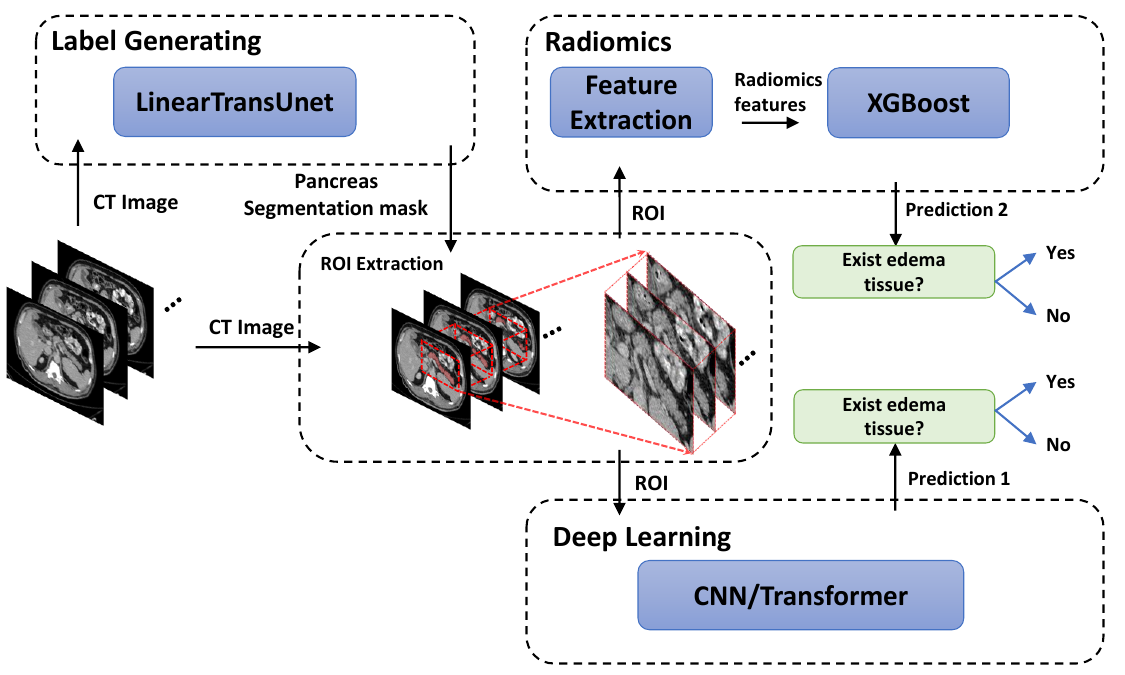}}
    \caption{Proposed workflow for detecting peri-pancreatic edema from CT images using two machine learning approaches: \textit{radiomics} and \textit{deep learning}. The process begins with inputting CT images into the \textit{LinearTransUnet}~\cite{zhang2022dynamic} segmentor for pancreas label generation. Segmented pancreases are in two complementary paths. In the first path, radiomics techniques are applied to the ROI (segmented pancreases) to extract features that are then used by an eXtreme Gradient Boosting (XGBoost) model to make a prediction (Prediction 2) regarding the presence of edema tissue. The second path utilizes the same ROIs with a deep learning-based model (CNN \&  Transformers) for diagnostic prediction. }
    \label{workflow}
\end{figure*}
\section{Introduction}
Pancreatitis poses a significant global healthcare concern due to its potential mortality risk~\cite{yao2023review,zhang2023deep}. The presence of peri-pancreatic edema serves as a crucial indicator of disease progression and prognosis. Thus, accurate peri-pancreatic edema detection and assessment are pivotal in early pancreatitis diagnosis and management. This is important because it can help determine the severity and progression of the condition, vital for improving patient prognosis. Despite advancements in the field of medical imaging with several innovative deep-learning approaches~\cite{carion2020end,touvron2023llama,jha2023transnetr,biswas2023adaptive}, the precise quantification and analysis of peri-pancreatic edema is still an unsolved problem. Thus, there is an emergent need for  developing techniques to solve the problem of peri-pancreatic edema detection.  

Deep learning is now a de-facto standard  to solve the challenges related to many medical imaging applications. However, applying these advanced deep learning (DL) techniques in the specific context of pancreatitis and peri-pancreatic edema detection has yet to be explored. As DL-based techniques could facilitate early diagnosis and treatment planning, this study aims to investigate the effectiveness of a DL-based segmentation algorithm, specifically \textit{LinTransUNet}, in accurately segmenting the pancreas from CT imaging data~\cite{zhang2022dynamic}. Subsequently, the extracted segmentation regions (ROI) are designed as training data for a Vision Transformer (ViT) model and radiomics-based XGBoost model in our experiments for peri-pancreatic edema prediction~\cite{dosovitskiy2020vit, NIPS2015_14bfa6bb}.
This research serves as a complementary tool to clinical diagnosis of pancreatitis by exploring the application of deep learning and radiomics models in pancreatic CT image analysis. Furthermore, the findings of this study hold valuable insights into the utilization of deep learning techniques for efficient medical data labeling. In summary, the contributions of this paper are as follows:

(1) We are not aware of any machine learning or deep learning methodology automatically detecting peri-pancreatic edema in CT scans. Therefore, our method serves as a strong baseline for peri-pancreatic edema detection even with conventional machine learning approaches. This is the first-ever application of pancreatic edema detection, to our best of knowledge.

(2) We provide a unique dataset for the public, comprising CT imaging data from 255 patients with detailed pancreas segmentation masks. This comprehensive dataset will serves as a valuable resource for the exploration and advancement of deep learning techniques in pancreatic research in general, offering a rich source of information for future research and clinical applications. The public links will be available upon acceptance of the paper.

(3) We introduce two distinct classifiers: a deep learning-based  and a radiomics-based classifier. The utility of the generated pancreas segmentation mask supports the radiomics analysis. These classifiers, built upon the extracted ROIs from our segmentation, serve as efficient tools for assisting medical professionals in diagnosing pancreatic diseases. Radiomics is a strong tool when the data is small. Herein, we combine the complementary power of both deep learning and radiomics to improve the detection of pancreatic peri-pancreatic edema. Deep learning and radiomics models as well as benchmarking methods will be available to public along with the data upon publication of the manuscript. 


\section{Methodology}
\textbf{Deep Learning Classification}. Once pancreases are segmented, regions tightly enclosing the pancreases can be used for automatic prediction of the peri-edema status. While we are using \textit{LinearTransUnet}~\cite{zhang2022dynamic} for the segmentation, one may use other segmentation strategies including semi-automated or manual labelings too. \textit{LinearTransUnet} is chosen as our segmentation method because of its outstanding performance in image segmentation.

We use a rich set of architecture family for benchmarking classification results. For convolutional neural networks (CNN) based comparison,  residual network (ResNet)~\cite{he2016deep}, densely connected convolutional networks (DenseNet)~\cite{huang2017densely}, MobileNetV3~\cite{howard2019searching}, MobileNetV2~\cite{sandler2018mobilenetv2}, and EfficientNet~\cite{dosovitskiy2020image} were included in the experiment. Representative transformer models like vision transformer (ViT)~\cite{dosovitskiy2020vit} and Swin transformer (Swin)~\cite{liu2021swin}~\cite{liu2022swin} were also analyzed for comprehensive study. Models are built strictly following their corresponding papers. The number of input channels is equal to 1 since input images are in grayscale. The trainable data was standardized to a consistent format and resized uniformly to $224 \times 224$ using linear interpolation. For deep learning models, we trained the model for 100 epochs in the batch size of 128 with 5-fold cross-validation, ensuring the robustness of the evaluations. Leveraging Stochastic Gradient Descent (SGD) with a momentum coefficient of 0.9 and original learning rate of $1e^{-5}$ as an optimizer, and cross-entropy as the loss function, respectively, the model aimed to optimize parameters and mitigate overfitting. 
The entire training process was executed using the PyTorch framework on an NVIDIA RTX A6000 GPU.

\textbf{Radiomics based classification}. In radiomics analysis, the necessity of the large quantities extraction of quantitative image features has been demonstrated in reports by Lambin~\cite{lambin2017radiomics} and Yao~\cite{yao2023review}. This study covers 107 distinct features extracted from CT images based on the generated masks. These features contain the following kinds: First Order Features, Shape Features, Gray Level Co-occurrence Matrix (GLCM) Features, Gray Level Size Zone Matrix (GLSZM) Features, Gray Level Run Length Matrix (GLRLM) Features, Neighbouring Gray Tone Difference Matrix (NGTDM) Features, Gray Level Dependence Matrix (GLDM) Features. XGboost was considered the key approach to build the radiomics-based classifier according to its outstanding performance~\cite{chen2016xgboost}. Optimal hyper-parameters were identified by adopting the grid search. This research set the following hyper-parameters: estimators = 3, maximum depth = 2.

\section{Experiments and Results}
\subsection{Dataset Overview}
We collected CT scans of 255 pancreatitis patients with IRB approval from Istanbul University Capa School of Medicine. There were two classes in the dataset: \textbf{179} pancreatitis patients \textbf{with peri-pancreatic edema} and \textbf{76} pancreatitis patients \textbf{without peri-pancreatic edema.} Each of the images had a corresponding standard pancreas mask and a label for disease class. Images with peri-pancreatic edema tissue were categorized as positive samples, denoted by label 1, other images are denoted with label 0. All standard pancreas segmentation masks were annotated by a team of expert radiologists (N=2) on consensus. After annotation, a careful evaluation was made to ensure the accuracy and reliability of the annotations. The standard pancreas segmentation masks were then used to evaluate the pancreas mask-generating model. Our dataset is available at \url{https://osf.io/wyth7/}.

\begin{figure}
\centering
\begin{minipage}[b]{0.32\linewidth}
\centering
\centerline{\includegraphics[width=\linewidth,height=0.75\linewidth]{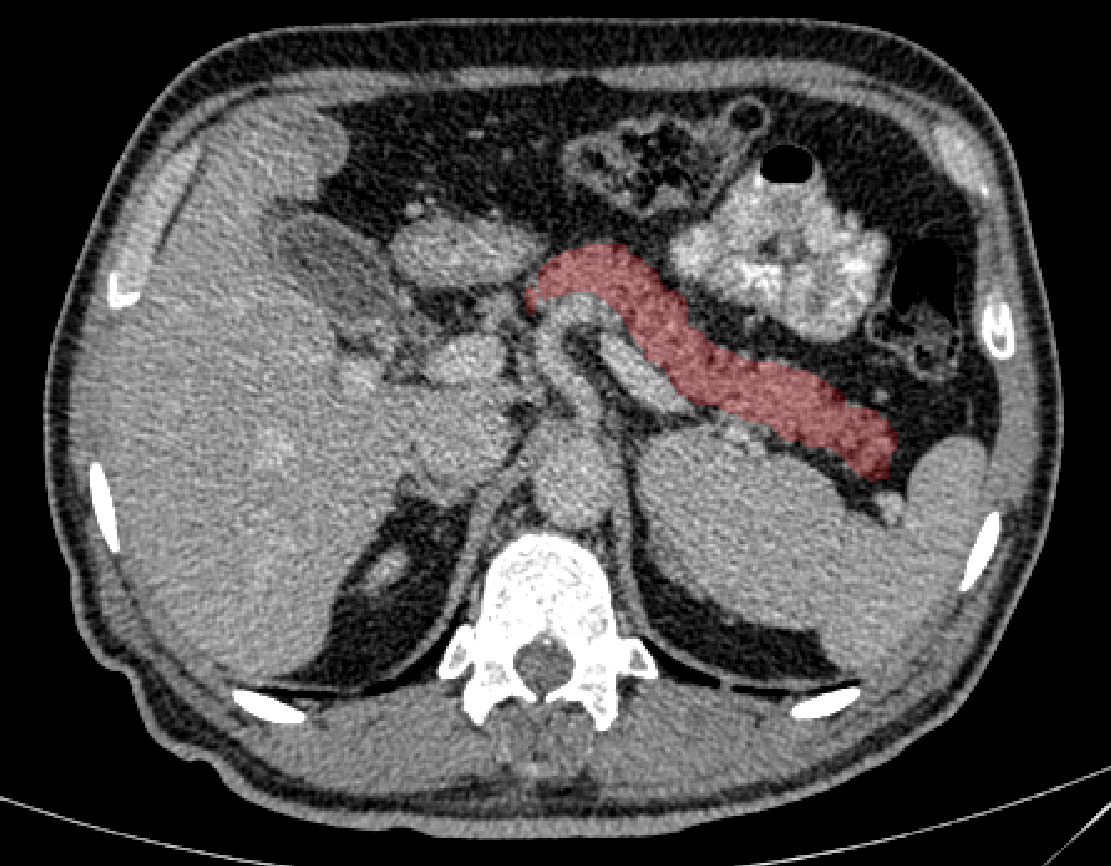}}
\vspace{0.05cm}
\end{minipage}
\begin{minipage}[b]{0.32\linewidth}
\centering
\centerline{\includegraphics[width=\linewidth,height=0.75\linewidth]{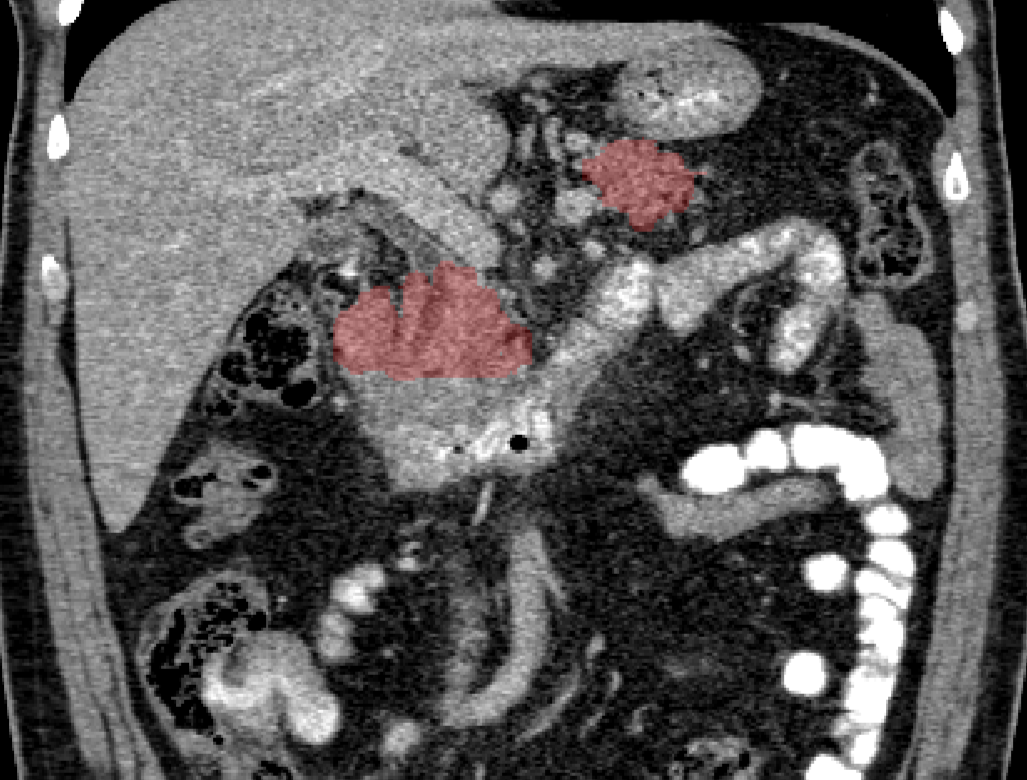}}
 \vspace{0.05cm}
\end{minipage}
\begin{minipage}[b]{0.32\linewidth}
  \centering
  \centerline{\includegraphics[width=\linewidth,height=0.75\linewidth]{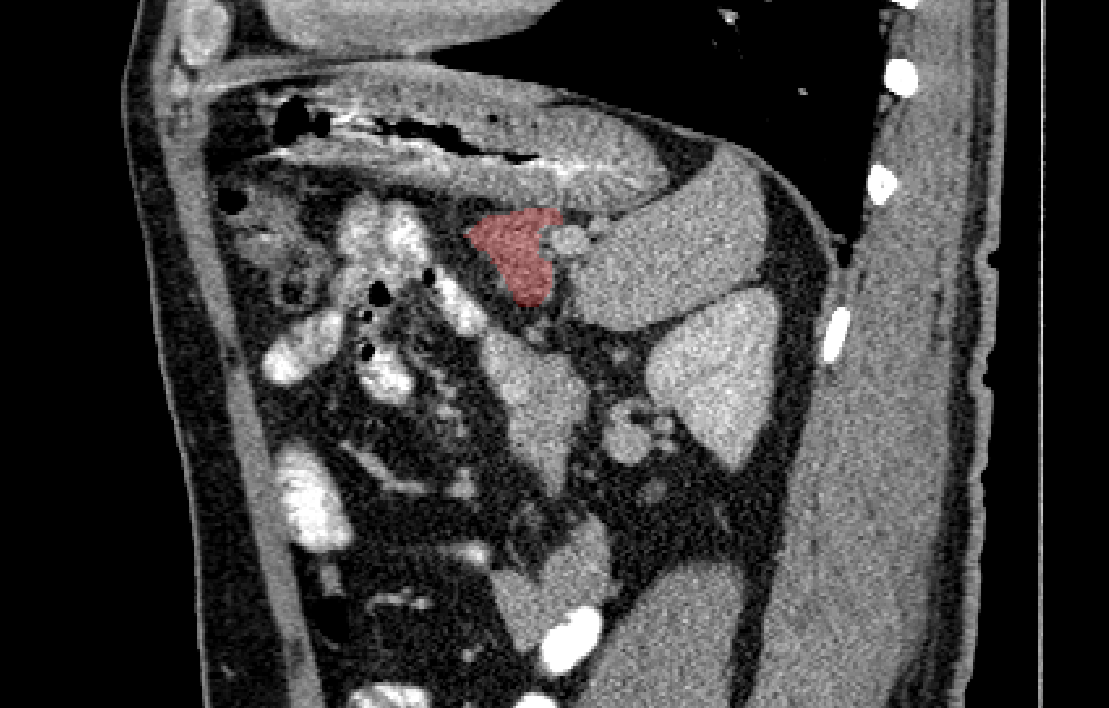}}
 \vspace{0.05cm}
\end{minipage}
\hfill
\begin{minipage}[b]{0.32\linewidth}
  \centering
  \centerline{\includegraphics[width=\linewidth,height=0.75\linewidth]{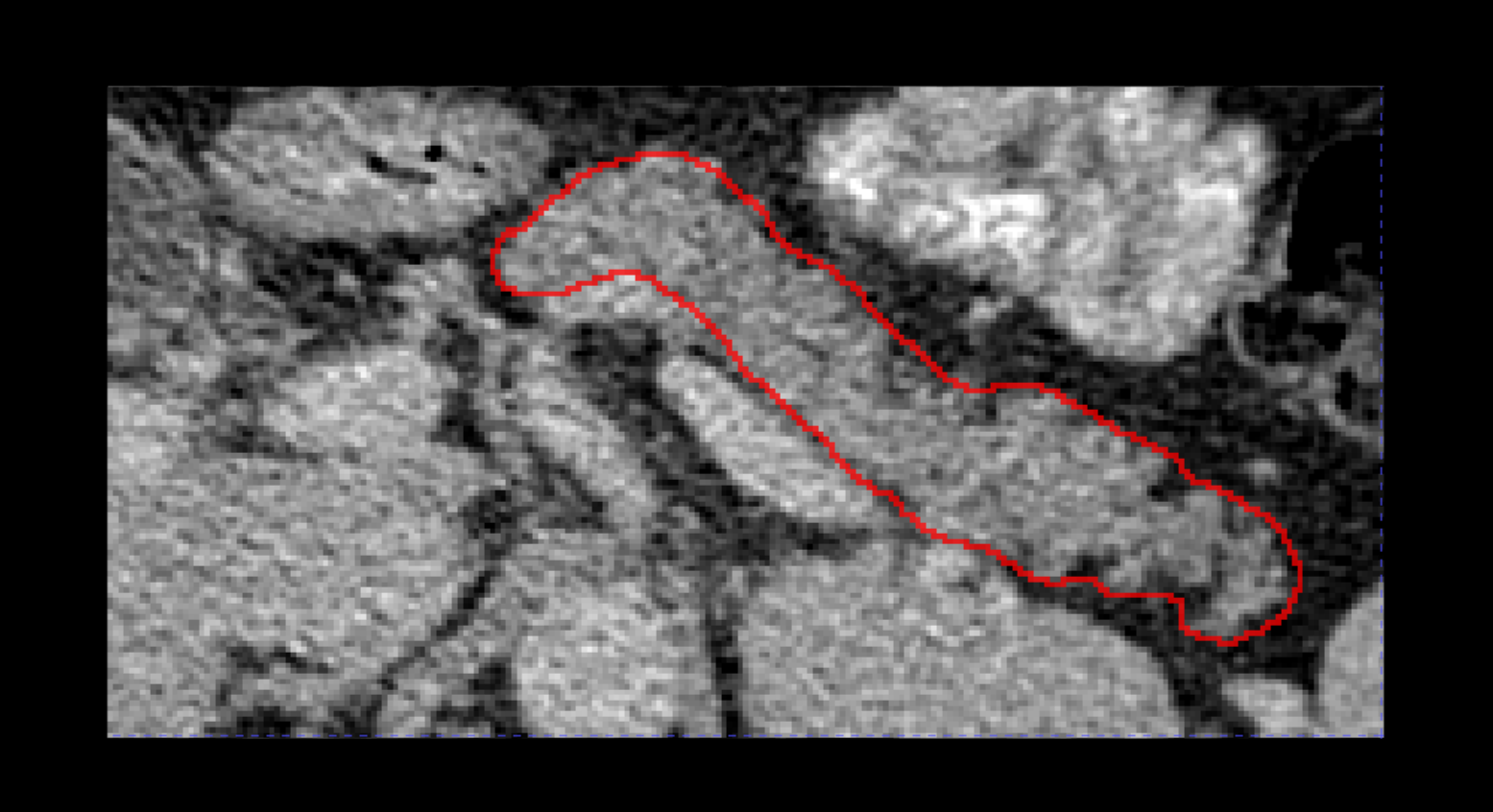}}
 \vspace{0.05cm}
\end{minipage}
\begin{minipage}[b]{0.32\linewidth}
  \centering
  \centerline{\includegraphics[width=\linewidth,height=0.75\linewidth]{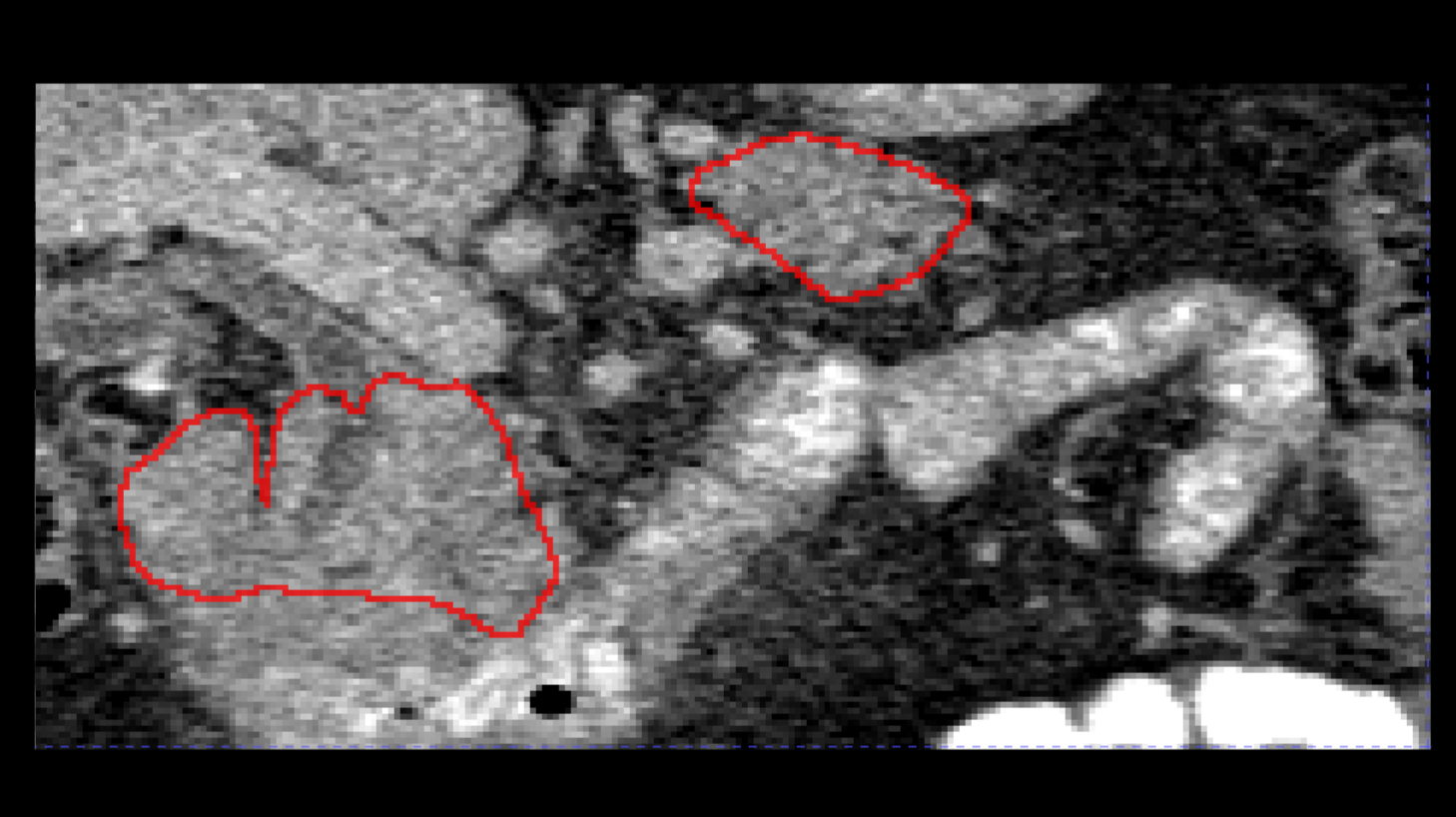}}
 \vspace{0.05cm}
\end{minipage}
\begin{minipage}[b]{0.32\linewidth}
  \centering
  \centerline{\includegraphics[width=\linewidth,height=0.75\linewidth]{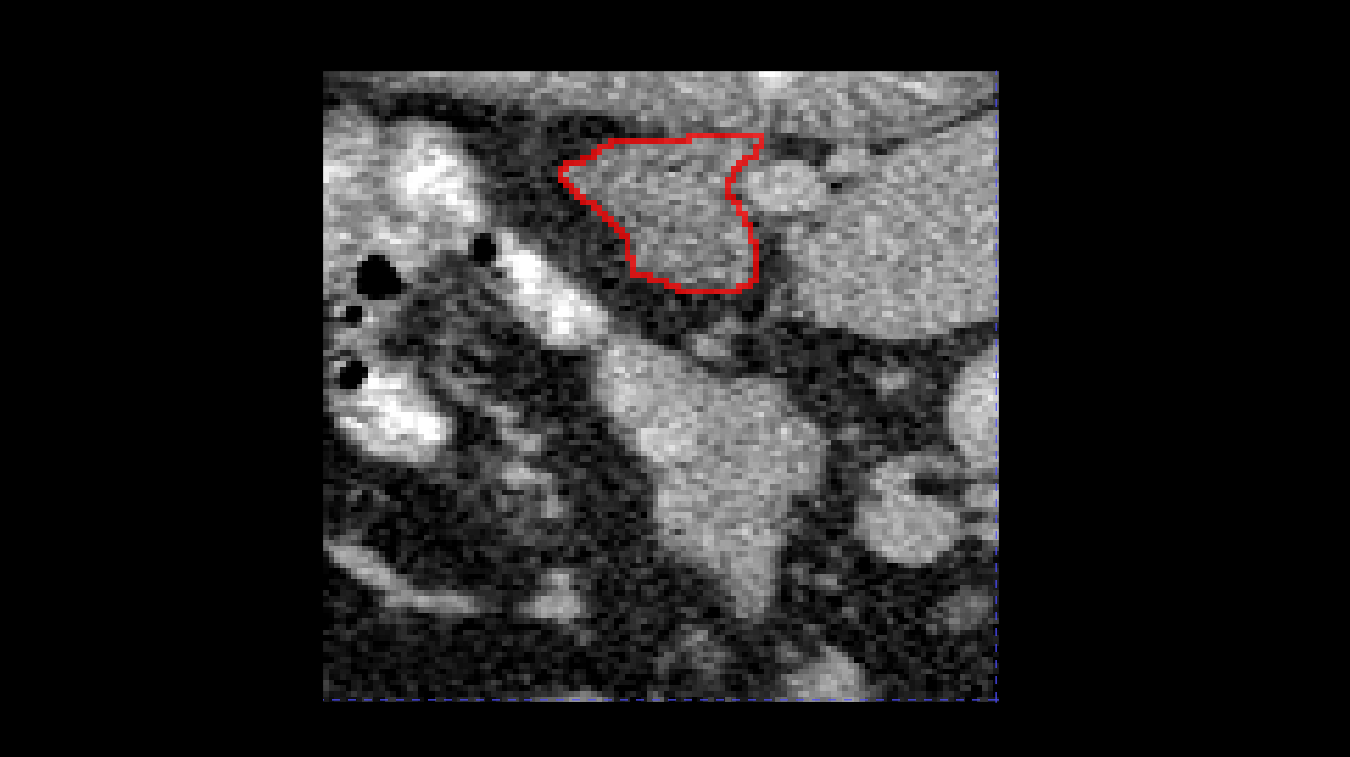}}
 \vspace{0.05cm}
\end{minipage}

    \caption{The upper row showcases the original CT scans, while the bottom row displays the extracted pancreas ROI. The images are presented in three different planes: axial, coronal, and sagittal, arranged from left to right.}
    \label{pancreas images and ROI}
\end{figure}

\subsection{Preprocessing}
The noise of medical instruments or motion blur can degrade the quality of medical images. Inspired by Yao's method~\cite{yao2023radiomics},  our first step was to perform reorientation. It is important, especially for cases where the reconstructed scan is not in the same orientation as with a mask. This is particularly important for the reproducibility of radiomics studies. Secondly, we applied conventional denoising algorithms (Gaussian denoising filter) to smooth out the noise. Once the denoising was performed, we used LinTransUNet~\cite{zhang2022dynamic} to segment the pancreases. 


We employed a small extension for the ROI enclosing the pancreas region. We used 10\% scaling in all axis to ensure the completeness of the pancreas region of over image. An extended ROI example is shown in Fig.~\ref{pancreas images and ROI}. 

\subsection{Segmentation Results}
For assessing the segmentation performance of LinTransUNet~\cite{zhang2022dynamic}, the Dice score, mIoU, precision, and recall were employed. With LinTransUNet, we obtained a Dice score of 80.85\%, mIoU of 68.73\%, precision of 76.21\%, and recall of 87.65\% as shown in Table~\ref{NN_performance}.



\begin{table}
\centering
\caption{Performance comparison of CNN and Transformer based approach for peri-pancreatic edema detection.}
\begin{tabular} {p{2.2cm}|p{1.56cm}|p{1.56cm}|p{1.56cm}} 
\toprule
{Backbone} & {Accuracy (\%)} & {Precision (\%)}   & {Recall (\%)} \\ \midrule
EfficientNet B0~\cite{dosovitskiy2020image}& 79.16 (±0.38) & 98.53 (±0.52)  &78.60 (±0.55)\  \\
MobileNetV2~\cite{sandler2018mobilenetv2}& 81.31 (±0.58) & 97.56 (±0.54)  & 80.97 (±0.72)  \\
MobileNetV3-s~\cite{howard2019searching} & 85.92 (±1.38) & 96.91 (±0.37)  & 85.73 (±1.40)  \\
ViT-tiny~\cite{dosovitskiy2020vit}    & 93.20 (±1.62) & 97.11 (±0.30)  &93.85 (±2.23)  \\
ResNet-50~\cite{he2016deep}    & 95.09 (±0.39) & 97.03 (±1.39)  &96.22 (±0.63)  \\ 
DenseNet-121~\cite{huang2017densely} & 96.41 (±0.34) & 98.21 (±0.61)  & 96.96 (±0.32)  \\
SwinV2-tiny~\cite{liu2022swin} & 97.27 (±0.52) & 98.60 (±0.80)  & 97.71 (±0.45)  \\
ResNet-34~\cite{he2016deep} & 97.63 (±0.20)  & 98.54 (±0.52)  & 98.25 (±0.43) \\
Swin-tiny~\cite{liu2021swin} & \textbf{97.95 (±0.40)} & \textbf{98.85 (±0.42)} & \textbf{98.38 (±0.17)} \\ 
\bottomrule
\end{tabular}
\label{NN_performance}
\end{table}

\begin{table}[!t]
\centering
\caption{Result of radiomics-based XGboost~\cite{chen2016xgboost} model on test data.}
\begin{tabular}{l|l|l|l}
\toprule

{Fold} & {Accuracy (\%)}& {Precision (\%)} & {Recall (\%)} \\  \midrule
 1    & 78.43                & 80.49 & 91.67 \\
 2    & 88.24                & 87.50& 97.22 \\
 3    & 76.47                & 80.00 & 88.89 \\
 4    & 78.43               & 82.05& 88.89\\
 5    & 76.47               & 79.49& 88.57 \\ 
\hline
\multicolumn{1}{l}{Average}  & 79.61 (± 4.04)   & 81.91 (± 2.93) & 91.05 (± 3.28) \\
\bottomrule
\end{tabular}
\label{Radoimics_performance}
\end{table}

\subsection{Classification}
The evaluation of the classifiers on our pancreas dataset relied on accuracy, precision, and recall as key metrics to provide further insights into the model's classifying of pancreatic and peri-pancreatic edema tissue. The outcomes obtained from 5-fold cross-validation are tabulated in Table~\ref{NN_performance}. As described in Table~\ref{NN_performance}, Swin-tiny owns the highest accuracy, precision, and recall. ResNet-34 also indicated a very high value in all metrics, and these values are close to those of Swin-tiny. SwinV2-tiny and DenseNet-121 also show relatively good results. The accuracy of the two models reached 96\%.
ResNet-50 and ViT-tiny underperform in the results, with 95\% and 93\% accuracy, respectively. The above models performed well in this experiment, with accuracy exceeding 90\%, precision exceeding 97\%, and recall exceeding 90\%. 

Table~\ref{NN_performance} demonstrates that MobilenetV2 MobileNetv3-s and EfficientNet B0 were relatively poor compared to others but still considerable in performance. MobilenetV2's accuracy was less than 86\%. Besides, its recall decreased below 86\%. A similar performance decrease was also shown in MobileNetv3-s and EfficientNet B0 and the recall drops to 78\% in the results of the EfficientNet B0, which indicates an erroneous diagnosis rate of nearly 22\% for patients with peri-pancreatic edema tissue. In contrast, the precision results were notably robust, each deep learning model exhibits a precision exceeding 96\%.  

With the same evaluation metrics as the deep learning model, the results of the radiomics-based XGBoost model are presented in Table~\ref{Radoimics_performance}. From this data, we can observe that the XGBoost model underperformed with both metrics compared with the deep learning model in general. XGBoost model has 79.61\% in accuracy, 81.91\% in precision, and 91.05\% in recall. Though the precision is 15\% lower than the worst-performed deep learning model, its low training time cost and fast processing speed still shows its potential as part of the auxiliary indicator.

\section{Further Analysis and Discussion}
Pancreas mis-segmentation can adversely affect classification results as it directly affects the ROI definition. For instance, regions distant from the pancreas could be misclassified as pancreatic tissue. Such a situation causes an unnecessary increase in size of the ROI. Going one step further, this issue could lead to the inaccuracy of feature extraction in radiomics and the increasing time in training the deep learning model. This issue warrants further investigation for potential remedies. For example, a post-processing scheme can be applied. In our experiments, we did not encounter such effect. 

Conversely, incomplete feature extraction from the pancreas and its surroundings on CT images results from the truncation issue. We empirically expanded the ROI area by 10\% to address this issue. Despite exhibiting considerable performance in terms of precision, the model faced challenges in recall, displaying a bias toward classifying images with peri-pancreas edema as normal pancreas tissue. In general, the performance difference of the traditional CNN model is relatively large in accuracy and recall. What can be seen in the deep learning model's performance is the model with a simpler structure tends to reach higher scores. In particular, though transformer models consume more time in training, all the transformer models gain very decent scores in all metrics. If further development of the model is required, the model that performs well in this experiment will be a great reference in building the model's structure. 

To further enhance the robustness of this experiment, it would be invaluable to explore variations in the model's performance by investigating different sizes for ROI expansion and diverse approaches to ROI extraction methodologies.

\section{Conclusion}
In this study, we established a very unique dataset comprising CT images from 255 pancreatic patients, categorized into classes indicative of the presence or absence of peri-pancreatic edema. Then, we employed LinTransUNet for the segmentation mask generation of CT pancreas images, yielding a commendable dice coefficient of 80.85\%. These generated segmentation masks were then utilized for ROI extraction, which is used for constructing two types of peri-pancreatic edema prediction models. We evaluated different deep learning backbones in this study. The best deep learning model (Swin-tiny) demonstrates robust accuracy rates averaging 97.95\%. While the radiomics-based model exhibited a lower score in every indicator, it achieved an accuracy of 79.61\% and a precision of 81.91\%. Furthermore, these segmentation masks were used to train the radiomics and deep learning-based classifier. By evaluating various deep learning and radiomics based models, we established a new baseline for peri-pancreatic edema detection.

\addtolength{\textheight}{-12cm}   

\bibliographystyle{plain}
\bibliography{main}
\end{document}